\documentclass[pra,twocolumn,showpacs,superscriptaddress]{revtex4}

\usepackage{amsmath,bm} 

\newcommand{\Exp}[1]{\,\mathrm{e}^{\mbox{\footnotesize$#1$}}}
\newcommand{\I}{\mathrm{i}}
\newcommand{\tr}{\mathop{\mathrm{Tr}}}

\begin{document}

\title{Symmetric construction of reference-frame-free qudits}
\author{Jun Suzuki}
\affiliation{Centre for Quantum Technologies, %
National University of Singapore, Singapore 117543, Singapore}
\affiliation{National Institute of Informatics, Chiyoda-ku,Tokyo 101-8430, Japan}
\author{Gelo Noel Macuja Tabia}
\affiliation{Centre for Quantum Technologies, %
National University of Singapore, Singapore 117543, Singapore}
\affiliation{Department of Physics, %
National University of Singapore, Singapore 117542, Singapore}
\author{Berthold-Georg Englert}
\affiliation{Centre for Quantum Technologies, %
National University of Singapore, Singapore 117543, Singapore}
\affiliation{Department of Physics, %
National University of Singapore, Singapore 117542, Singapore}

\date{15 August 2008}

\begin{abstract}
By exploiting a symmetric scheme for coupling $N$ spin-1/2 constituents
(the physical qubits) to states with total angular momentum $N/2-1$, we
construct rotationally invariant logical qudits of dimension $d=N-1$.
One can encode all qudit states, and realize all qudit measurements,
by this construction.
The rotational invariance of all relevant objects enables one to transmit
quantum information without having aligned reference frames between the
parties that exchange the qudits.
We illustrate the method by explicit constructions of reference-frame-free
qubits and qutrits and, for the qubit case, comment on possible experimental
implementations. 
\end{abstract}

\maketitle

%
\section{Introduction}
The raw experimental data about physical systems or events are almost always
tied to reference frames, defined by the coordinate systems to which the data
refer.
The comparison of data acquired by different observers then requires that they
know how their reference frames are related to each other, such as whether the
axes of the coordinate systems are aligned or rotated.

In the context of quantum information theory, the role of the reference frame
has been reconsidered recently; 
see \cite{BRS} for a summary.
There is, in particular, an intimate connection with the concept of
decoherence free (DF) subsystems and subspaces, 
which are important for experimental implementations of schemes for processing
quantum information.
In quantum cryptography, for example, the presence of decoherence 
lowers the efficiency of the quantum channel involved. 
Moreover, the lack of a shared reference frame between two distant parties 
becomes a practical problem when establishing a secure channel between them. 

It is, therefore, reasonable to ask whether or not one could be free, in
general, from the problem of decoherence or sharing of the reference frame.
Put differently, one may ask if it is possible to construct arbitrary
quantum states which are DF or reference-frame-free (RFF). 
This question has been widely discussed \cite{BRS}, and general arguments
ensure the existence of such DF subsystems, DF subspaces, and RFF quantum
states in any finite dimension.  
When it comes to their explicit construction, however, one
encounters a situation in which one has to work out details whose number
increases exponentially with the dimension of the Hilbert space in question.

In this contribution, we narrow this gap between the in-principle
possibility and the in-practice difficulty by 
an explicit construction of all states of a 
$d$-dimensional RFF quantum system, the \textit{RFF qudit}, 
out of $N=d+1$ spin-1/2 constituents.
We are thus making \emph{logical} RFF qudits out of \emph{physical} qubits,
and the construction also identifies the $d$-dimensional DF subsystem. 

The standard construction is based on successive addition of the angular
momenta of the spin-1/2 ingredients with Clebsch-Gordan coefficients for the
probability amplitudes. 
With more than two constituents, the classification of the final states
depends on the order in which the individual spins are added, and the
complexity grows very rapidly with the number of constituents \cite{AMinQP}.

Alternatively, there is the symmetric coupling that we will exploit here
\cite{sym1,sym2}. 
The general symmetric coupling is known only for three angular momenta so far,
and it is 
presently unknown if there is a symmetric coupling scheme
for more than three angular momenta. 
In the special case of the coupling of $N$ identical angular momenta, our
study suggests the possibility of the symmetric coupling. 
In this paper we report the $N$ spin-1/2 case and, taking advantage of the
symmetric coupling scheme, we construct the RFF qudit immediately without the
need for evaluating the Clebsch-Gordan coefficients. 
 
The paper is organized as follows. 
We first present the symmetric coupling of $N$ spin-1/2 constituents for
the states with second-largest total angular momentum in Sec.~II.  
We construct the general RFF qudit in Sec.~III, and 
illustrate the procedure for RFF qubits and RFF qutrits in Sec.~IV.  

%
\section{Symmetric coupling of $N$ spin-1/2 constituents}
The direct product of $N$ spin-1/2 systems is a direct sum of irreducible
representations of angular momentum,  
\begin{equation} \label{decom}
\mathcal{D}_{1/2}^{\otimes N}=\bigoplus_{j \in J}c_j\mathcal{D}_j,
\end{equation}
where the index set $J=\{N/2,N/2-1,\dots\}$ has $(N+1)/2$ elements if $N$ is
odd, and has $N/2+1$ elements if $N$ is even.  
Here,
\begin{equation}
c_j=\frac{N!\,(2j+1)}{(N/2+j+1)!\,(N/2-j)!}  
\end{equation}
is the multiplicity of $\mathcal{D}_j$, the irreducible representation of
angular momentum $j$. 
Our main concern is the subsystem of the second-largest angular momentum
states, where we identify the $d$-dimensional DF subsystem or RFF qudit.

The states with second-largest angular momentum, ${j_2=N/2-1}$, 
have a multiplicity of ${c_{j_2}=N-1}$.  
Therefore, these states can be labeled by the eigenvalues of the $z$-component
of the total angular momentum  
${\vec{J}=\sum_{\ell=1}^{N}\vec{\sigma}^{(\ell)}/2}$ together with the
degeneracy label $\lambda$.  
That is, the state kets are denoted by $|j_2,m_2;\lambda\rangle$ with 
\begin{equation}
m_2=j_2,j_2-1,\dots, -j_2\  
\mathrm{and}\ \lambda=1,2,\dots, N-1\,.
\end{equation}
There are $d=2j_2+1=N-1=c_{j_2}$  states for each value of $\lambda$, so that
we have $d^2=(N-1)^2$ states for $j_2=N/2-1$ in total. 

Upon denoting the kets for the single--spin-1/2 states with $m=1/2$ 
and $m=-1/2$ by $|0\rangle$ and $|1\rangle$, respectively, we have 
$|0^{\otimes N}\rangle=|00 \dots 0\rangle$ for the unique state with maximal
values of both $j$ and $m$, that is $j_1=m_1=N/2$, and multiple applications
of the ladder operator $J_-={J_x-\I J_y}$ yield all states of maximal total
angular momentum in the familiar way, 
$|j_1,m_1\rangle\propto J_{-}^{j_1-m_1}|0^{\otimes N}\rangle$.
Supplementing $J_-$ are its $d$ orthogonal partners $\Omega_{-}(\lambda)$,
defined by
\begin{equation} \label{omega}
\Omega_{-}(\lambda)=\frac{1}{\sqrt{N}}\sum_{\ell=1}^{N}
\omega_{N}^{\lambda\ell} \sigma_{-}^{(\ell)}
\qquad\mbox{with $\omega_{N}=\Exp{2\pi\I/N}$},
\end{equation}
where $\sigma_{-}^{(\ell)}$ is the lowering operator for the $\ell$th
constituent.  
The angular momentum states with $m_2=j_2$ are then obtained as 
$|j_2,j_2;\lambda\rangle=\Omega_{-}(\lambda)|0^{\otimes N}\rangle$, and
successive applications of $J_{-}$ give the remaining
$|j_2,m_2;\lambda\rangle$.  
Since $\Omega_{-}(\lambda)$ and $J_{-}$ commute with each other, we have, 
\begin{align} \label{symmcoup}\nonumber
|j_2,m_2;\lambda\rangle 
&=\sqrt{\frac{(j_2+m_2)!}{(2j_2)!(j_2-m_2)!}}\,
\Omega_-(\lambda)J_-^{j_2-m_2}|0^{\otimes N}\rangle \\[1ex] 
& \propto \Omega_-(\lambda) |j_1,m_2+1\rangle \,,
\end{align}
for which
$\langle j_2,m_2^{\,};\lambda|j_2,m_2';\lambda'\rangle
=\delta_{{m_2^{\,}m_2'}}\delta_{\lambda\lambda'}$ states their orthonormality. 

The discrete Fourier transformation that we chose in
(\ref{omega}) is just one possibility for defining the  
$\Omega_-(\lambda)$s and thus the kets $|j_2,m_2;\lambda\rangle$. 
More generally, any unitary $d\times d$ matrix $U$, with $N$th-row matrix
elements $U_{N\ell}=N^{-1/2}$, can serve in 
$\Omega_{-}(\lambda)=\sum_{\ell=1}^{N} U_{\lambda \ell}\sigma_{-}^{(\ell)}$. 
For the specific choice of the discrete Fourier matrix, the projectors 
$|j_2,m_2;\lambda\rangle\langle j_2,m_2;\lambda|$ are invariant 
under the cyclic permutation $\vec{\sigma}^{(1)}\to\vec{\sigma}^{(2)}%
\to\cdots\to\vec{\sigma}^{(N)}\to\vec{\sigma}^{(1)}$ 
of the spin-1/2 constituents.  
An analogous construction works for systems 
of $N$ constituents with spin other than $1/2$.

Regarding the permutation symmetry, we note that the
unitary permutation operators are
invariant when the same unitary transformation is applied to all constituents,
and therefore the permutation operators can be composed to form invariant
operators.  
By requiring that the invariant objects are hermitian and nonnegative, one can
then construct a proper quantum state.  
This fact was utilized in constructing the Werner state for two parties
\cite{werner}, and recently generalized to more parties~\cite{EW,CK}.

%
\section{Reference-frame-free qudit}
The basic operators for the RFF qudit in $d=N/2-1=2j_2+1$ dimension are the
$d^2$ operators 
\begin{equation} \label{Qoperator}
Q_{\lambda\lambda'}=\sum_{m_2=-j_2}^{j_2}|j_2,m_2;\lambda\rangle\langle
j_2,m_2;\lambda' |=Q_{\lambda'\lambda}^{\dagger}\,. 
\end{equation}
They commute with the vector operator of total angular momentum, 
$\vec{J} Q_{\lambda\lambda'}=Q_{\lambda\lambda'}\vec{J}$, 
and are closed under multiplication, 
$Q_{\lambda\lambda'}Q_{\lambda''\lambda'''}=%
\delta_{\lambda'\lambda''}Q_{\lambda\lambda'''}$. 
It follows that $Q_{\lambda\lambda'}$ can be written in the tensor product form 
\begin{equation}
Q_{\lambda\lambda'}=I_d \otimes |\lambda\rangle\langle\lambda'| \,,
\end{equation}
where the first factor refers to the \textit{signal} qudit, 
and the $d$-dimensional identity $I_d$ refers to the \textit{idler} qudit. 

This signal-idler split is reminiscent of the split into visible and
hidden degrees of freedom in Ref.~\cite{Adamson+al}, yet these are different
splits.
The indistinguishability of the constituents, central to the 
visible-hidden split, does not interfere with the signal-idler split because
we take for granted that the constituents are in different spatial modes. 

A qudit quantum state, specified by a $d\times d$ density matrix 
$\rho=\rho^{\dagger}\ge 0$ with matrix elements $\rho_{\lambda\lambda'}$, 
is then implemented by 
\begin{equation}
\rho^{\mathrm{(RFF)}}=\frac{1}{d}I_d\otimes\sum_{\lambda,\lambda'=1}^{d}
|\lambda\rangle\rho_{\lambda\lambda'}\langle\lambda'|\,, 
\end{equation}
where the signal qudit is in the state $\rho$ and the idler qudit is in the
completely mixed state. 
Whereas $\rho^{\mathrm{(RFF)}}$ is a mixed state of the $N$ spin-1/2
constituents, with a binary entropy of 
$S(\rho^{\mathrm{(RFF)}})=S(\rho)+\log_2 d$, the state of the signal qudit can
be pure or mixed, whatever is the nature of the given qudit state $\rho$. 

Any qudit positive-operator-valued measure (POVM), $\sum_k \Pi_k=I_d$ with 
$\Pi_k=\Pi_k^{\dagger}\ge0$, can be realized as a POVM for 
the signal qudit by the analogous construction 
\begin{equation}
\Pi_k^{(\mathrm{RFF})}=I_d \otimes\sum_{\lambda,\lambda'=1}^{d}
 |\lambda\rangle(\Pi_k)_{\lambda\lambda'}\langle\lambda'|\,,
\end{equation}
so that $\tr \{\rho^{\mathrm{(RFF)}}\Pi_k^{(\mathrm{RFF})} \}=%
\tr\{ \rho\Pi_k \}$ for all outcomes $\Pi_k$ of the POVM under consideration. 

The ambiguity in defining $\Omega_-(\lambda)$, mentioned in the paragraph
following Eq.~(\ref{symmcoup}), carries over to $\rho^{(\mathrm{RFF})}$ and
$\Pi_k^{(\mathrm{RFF})}$, which are equally ambiguous. 
But once the $\Omega_-(\lambda)$ are chosen, the above construction gives a
definite implementation of $\rho$ and $\Pi_k$.  

\section{Examples}
%
\subsection{Theoretical construction of the RFF qubit ($N=3$)}
As a first illustration we consider the RFF qubit (${d=2}$, ${j_2=1/2}$)
composed of ${N=3}$ spin-1/2 constituents.  
It is both convenient and systematic to express all operators in terms of the
unitary and hermitian swap operators $P_{jk}=P_{kj}=(1+\vec{\sigma}^{(j)}\cdot
\vec{\sigma}^{(k)})/2$ (for $j\neq k$) that permute the $j$th and $k$th
constituents: 
$P_{jk} \vec{\sigma}^{(k)}=\vec{\sigma}^{(j)}P_{jk}$.  
Since these swap operators are obviously invariant under the collective
rotations, so are
\begin{equation}\label{RFFqubit-1}
Q_{12}=Q_{21}^{\dagger}=\frac13 (P_{12}+\omega_3P_{23}+\omega_3^2 P_{31})\,, 
\end{equation}
and 
\begin{align}\label{RFFqubit-2}
\left.\begin{array}{c@{}}
Q_{11}=Q_{11}^\dagger=Q_{12}Q_{21}\\[1ex]
Q_{22}=Q_{22}^\dagger=Q_{21}Q_{12}
\end{array}\right\}
=&\frac{1}{2}-\frac{1}{6}(P_{12}+P_{23}+P_{31})
\nonumber\\&\mp\frac{\I}{\sqrt{12}}[P_{31},P_{12}]\,.
\end{align}

The components of the hermitian Pauli vector $\vec{\sigma}^{(\mathrm{RFF})}$
for the RFF qubit are then given by \cite{filippo}  
\begin{align}
  \label{RFFqubit-3} \nonumber
  \sigma_x^{(\mathrm{RFF})}&=Q_{12}+Q_{21}=\frac{1}{3}(2P_{12}-P_{23}-P_{31})
\nonumber\\&=
  \frac{1}{6}\bigl(2\vec{\sigma}^{(1)}\cdot\vec{\sigma}^{(2)}
-\vec{\sigma}^{(2)}\cdot\vec{\sigma}^{(3)}
-\vec{\sigma}^{(3)}\cdot\vec{\sigma}^{(1)} \bigr)\,, \nonumber\\
  \sigma_y^{(\mathrm{RFF})}&=-\I Q_{12}+\I Q_{21}
=\frac{1}{\sqrt{3}}(P_{23}-P_{31})\nonumber\\&
=\frac{1}{\sqrt{12}}\bigl(\vec{\sigma}^{(2)}\cdot\vec{\sigma}^{(3)}
-\vec{\sigma}^{(3)}\cdot\vec{\sigma}^{(1)}\bigr)\,,\nonumber\\
  \sigma_z^{(\mathrm{RFF})}&=Q_{11}-Q_{22}=\frac{-\I}{\sqrt{3}}[P_{31},P_{12}]
\nonumber\\&=
-\frac{1}{\sqrt{12}}\bigl(\vec{\sigma}^{(1)}\times\vec{\sigma}^{(2)}\bigr)
\cdot\vec{\sigma}^{(3)}\,,
\end{align}
and the projector onto the ${j_2=1/2}$ subspace of the signal and idler qubits is 
\begin{equation}
I_{j=1/2}=Q_{11}+Q_{22}=1-\frac{1}{3}(P_{12}+P_{23}+P_{31})\,.
\end{equation} 
The explicit expressions above emphasize once more that these operators are rotationally invariant, indeed,
and therefore any orientation in space of the $x,y,$ and $z$ axes of the
Cartesian reference frame is as good as any other.  

Here is an example of a state preparation.
By preparing the third spin-1/2 constituent in a completely mixed state and
the first and second in their singlet state, the experimenter puts the three
physical qubits into the mixed state 
\begin{align}
\rho_3&=\frac{1}{4}\bigl(1-\vec{\sigma}^{(1)}\cdot\vec{\sigma}^{(2)}\bigr)
=\frac{1}{2}(1-P_{12})
\nonumber\\
&=\frac{1}{2} (Q_{11}+Q_{22}-Q_{12}-Q_{21})
\mathrel{\widehat{=}}
\frac{1}{2}\begin{pmatrix}1&-1\\-1&1\end{pmatrix},
\end{align}
which is a pure state for the signal qubit. Likewise, $\rho_1=(1-P_{23})/2$
and $\rho_{2}=(1-P_{31})/2$ are pure states of the signal qubit, and together
these states make up a trine. 
Such trine states could, for instance, be used for an implementation of secure
quantum key distribution \cite{trine}. 

\subsection{Remarks on experimental implementation}
If photon polarization is used for the realization of the spin-1/2
constituents, one could begin with an entangled pure four-photon state and
measure the fourth photon in a suitable way to put the other three photons
into polarization trine states.
In addition, the three trine states make up a POVM,
\begin{align}
I_{j=1/2}&=
  \frac{1}{6}\bigl(3-\vec{\sigma}^{(1)}\cdot\vec{\sigma}^{(2)}
                    -\vec{\sigma}^{(2)}\cdot\vec{\sigma}^{(3)}
                    -\vec{\sigma}^{(3)}\cdot\vec{\sigma}^{(1)}\bigr)
\nonumber\\
   &=  \frac{2}{3}(\rho_1+\rho_2+\rho_3)\,,
\end{align}
and a possible optical implementation of this POVM is described in
Ref.~\cite{Janos}. 
The basic ingredient is the conversion, with the aid of a quantum
teleportation protocol, of the original three-photon polarization analog of
the three spin-1/2 system into a single-photon analog. 
When thus having two spin-1/2 analogs in spatial degrees of freedom 
of the photon and the third in the polarization degree of freedom,  
it is rather straightforward to implement any POVM using phase
shifters, beam splitters, and photon counters only.  

Another experimental realization, for the purposes of quantum storage, would
make use of a trio of neutral spin-1/2 atoms, trapped in an optical lattice,
at a distance from the nearest other trio. 
Qubits stored as the signal qubits of such trios are protected against stray
magnetic fields of arbitrary, but equal strength at the three lattice sites of
one trio.
The details of an actual implementation of such a quantum memory device are
presently being investigated. 

%
\subsection{General case}
When one considers the higher-dimensional general case, it is useful to employ
a more systematic construction of the RFF qudit. 
One possible way is to expand the state in terms of the generators for a
SU$(d)$ Lie group with real coefficients. 
The standard Gell-Mann matrices together with the positivity requirement
provide proper $d$ level quantum states \cite{kimura,bryd}. 
Alternatively, one can use the unitary Heisenberg-Weyl-Schwinger (HWS)
operator basis \cite{schwinger}. 
The complete set of unitary operators are given by
$U^jV^k$($j,k=1,2,\dots,d$), where the unitary operators $U$ and $V$ have
period $d$, $U^d=V^d=1$, and $U^jV^k=\omega_d^{-jk}V^kU^j$ with
$\omega_d=\exp(2\pi \I/d)$.  

Explicitly, the RFF $U$ and $V$ unitary operators are given by 
\begin{equation}\label{HWS}
U_{d}=\sum_{\lambda=1}^{d}
\omega_d^{\lambda}Q_{\lambda\lambda}\,,\quad 
V_{d}=\sum_{\lambda=1}^{d-1}Q_{\lambda\lambda+1}+Q_{d1}\,.
\end{equation} 
For the RFF qubit of Eqs.~(\ref{RFFqubit-1})--(\ref{RFFqubit-3}), these are 
simply ${U_2=-\sigma_z^{(\mathrm{RFF})}}$ and ${V_2=\sigma_x^{(\mathrm{RFF})}}$. 

%
\subsection{Theoretical construction of the RFF qutrit ($N=4$)}
The next example is the case of four spin-1/2 constituents. This system is
known to provide the DF qubit subspace spanned by two spin-$0$ states, 
which have
been studied using the standard Clebsch-Gordan coefficients, and have been
experimentally demonstrated \cite{DFSexperiment}.  
Our emphasis here is the usefulness of the RFF qutrit construction. 
As a byproduct, the symmetric coupling yields an alternative choice
for the basic spin-$0$ states, with properties different from the spin-0
states obtained by the Clebsch-Gordan coupling scheme.  

For ${N=4}$, the second-largest angular momentum states have ${j_2=1}$ and are
triply degenerate. 
With ${\omega_4=\I}$, they are   
\begin{align}
|1,1;\lambda\rangle&=\frac{1}{2}\bigl(
\omega_4^{\lambda}|1000\rangle+\omega_4^{2\lambda}|0100\rangle\nonumber\\
&\hphantom{=\frac{1}{2}\bigl(}
+\omega_4^{3\lambda}|0010\rangle+|0001\rangle\bigr)\,,\nonumber\\ 
|1,0;\lambda\rangle&=\frac{1}{\sqrt{8}}
\bigl[(\omega_4^{\lambda}+1)\bigl(|1001\rangle-|0110\rangle\bigr)\nonumber\\
&\hphantom{=\frac{1}{\sqrt{8}}\bigl[}
+(\omega_4^{2\lambda}+1)\bigl(|0101\rangle-|1010\rangle\bigr)\nonumber\\
&\hphantom{=\frac{1}{\sqrt{8}}\bigl[}
+(\omega_4^{3\lambda}+1)\bigl(|0011\rangle-|1100\rangle\bigr)\bigr]\,,
\nonumber\\ 
|1,-1;\lambda\rangle&=-\frac{1}{2}\bigl(
\omega_4^{\lambda}|0111\rangle+\omega_4^{2\lambda}|1011\rangle\nonumber\\
&\hphantom{=-\frac{1}{2}\bigl(}
+\omega_4^{3\lambda}|1101\rangle+|1110\rangle\bigr)\,,
\end{align}
when expressed as superpositions of the basic product states of the four
spin-1/2 constituents. 

We introduce the following hermitian operators for convenience. 
\begin{align} \nonumber
A_1&=P_{12}\!-\!P_{34},\ A_2=P_{13}\!-\!P_{24},\ A_3=P_{14}\!-\!P_{23},\\ \nonumber
K_1&=\I [P_{23},P_{24}],\quad K_2=\I [P_{34},P_{13}],\\ \nonumber
K_3&=\I [P_{14},P_{24}],\quad K_4=\I [P_{12},P_{13}],\\
L_1&=P_{12}P_{34},\ L_2=P_{13}P_{24},\ L_3=P_{14}P_{23}.
\end{align}
With these notations, the basic operators (\ref{Qoperator}) defined above are 
\begin{align}  \nonumber
Q_{11}&=\frac{1}{4}\bigl[1-\frac12(K_1+K_2+K_3+K_4)-L_2 \bigr],\\ \nonumber
Q_{22}&=\frac{1}{4}(1-L_1+L_2-L_3),\\ \nonumber
Q_{33}&=\frac{1}{4}\bigl[1+\frac12(K_1+K_2+K_3+K_4)-L_2 \bigr],\\ \nonumber
Q_{12}&=\frac{1}{8}\bigl[(1\!+\!\I)A_1-(1\!-\!\I)A_3-\I(K_1\!-\!K_3)-(K_2\!-\!K_4) \bigr]\\ \nonumber
&=Q_{21}^{\dagger},\\ \nonumber
Q_{23}&=\frac{1}{8}\bigl[(1\!+\!\I)A_1-(1\!-\!\I)A_3+\I(K_1\!-\!K_3)+(K_2\!-\!K_4) \bigr]\\ \nonumber
&=Q_{32}^{\dagger},\\
Q_{13}&=\frac{1}{4}\bigl[A_2+\I(L_1-L_3) \bigr]=Q_{31}^{\dagger}.
\end{align}
And the construction (\ref{HWS}) gives
\begin{align}
U_3&=\frac{\omega_3^2}{4}\bigl[ -L_1+2L_2-L_3+\sqrt{3}\I(K_1\!+\!K_2\!+\!K_3\!+\!K_4) \bigr]\,,\nonumber\\
V_3&=\frac{1}{4}\bigl[\I(L_1-L_3)+(1+\I)A_1+A_2-(1-\I)A_3) \bigr]
\end{align}
for the basic HWS unitary operators. 

Similarly, the $j_3=0$ states are given by
\begin{align} \label{symsing}
|0,0;\lambda\rangle=\frac{1}{\sqrt{6}}\bigl[&
\omega_3^{\lambda}\bigl(|1001\rangle+|0110\rangle\bigr) \nonumber\\
&+\omega_3^{2\lambda}\bigl(|0101\rangle +|1010\rangle\bigr)\nonumber\\
&+\bigl(|0011\rangle+|1100\rangle\bigr)\bigr]\,. 
\end{align}
Quite obviously, the two respective projectors ($\lambda=1,2$)  
\begin{align}
|0,0;\lambda\rangle\langle 0,0;\lambda|
=&\frac{1}{3}(S_{12}S_{34}+S_{13}S_{24}+S_{14}S_{23})\nonumber\\
&+\frac{\I(-1)^{\lambda}}{\sqrt{12}}\bigl(
[S_{12},S_{13}]-[S_{23},S_{24}]\\\nonumber
&\hphantom{+\frac{\I(-1)^{\lambda}}{\sqrt{12}}\bigl(}
+[S_{34},S_{31}]-[S_{41},S_{42}]\bigr)
\end{align}
are unchanged or interchanged when the spin-1/2 constituents are permuted. 
Here, $S_{jk}=(1-\vec{\sigma}^{(j)}\cdot \vec{\sigma}^{(k)})/4$ is the singlet
state between $j$th and $k$th constituents. 
This invariance is in marked contrast to the lack of permutation invariance in
the projectors  
\begin{align} 
|S_1\rangle\langle S_1|&=\frac13
(-S_{12}S_{34}+2S_{13}S_{24}+2S_{14}S_{23})\,,
\nonumber\\ \label{CGproj} 
|S_2\rangle\langle S_2|&=S_{12}S_{34}
\end{align}
onto the $j_3=0$ states of the standard successive coupling,
\begin{align} \label{CGsing}
|S_1\rangle&=\frac{1}{2}\bigl(|0101\rangle+|1010\rangle
-|1001\rangle-|0110\rangle\bigr)\,,\nonumber\\
|S_2\rangle&=\frac{1}{\sqrt{12}}\bigl(2|0011\rangle+2|1100\rangle-|0101\rangle
\nonumber\\&\hphantom{=\frac{1}{\sqrt{12}}\bigl(}
-|1010\rangle-|0110\rangle-|1001\rangle\bigr)\,.
\end{align}

We note that the singlet states (\ref{symsing}) have been introduced in
Ref.~\cite{HS}.  
A detailed comparison of the properties of the states (\ref{symsing}) and the
states (\ref{CGsing}) establishes that they should be regarded as quite
different singlet bases \cite{HS,BH}.  
Put differently, the total angular momentum and its third component together
with the degeneracy do not specify the state uniquely; one needs other quantum
numbers in addition. 
In the case of $N$ qubits, for example, the product states of the 
computational basis  are uniquely labeled by $N$ binary quantum numbers. 
Although this ``missing label problem'' is well known \cite{peccia}, 
it seems that it has not received due attention  
in quantum information theory as yet.  

Lastly, for completeness we report the RFF qubit Pauli operators for the
$j_3=0$ qubit.
They are
\begin{align} 
\sigma_x^{(\mathrm{RFF})}&=\frac{-2}{3}(2S_{12}S_{34}-S_{14}S_{23}-S_{13}S_{24})
\,,\nonumber\\ 
\sigma_y^{(\mathrm{RFF})}&=\frac{-2}{\sqrt{3}}(S_{13}S_{24}-S_{14}S_{23})\,,
\nonumber\\
\sigma_z^{(\mathrm{RFF})}&=\frac{-\I}{\sqrt{3}}\bigl(
[S_{12},S_{13}]-[S_{23},S_{24}]\nonumber\\
&\hphantom{=\frac{-\I}{\sqrt{3}}\bigl(}
+[S_{34},S_{31}]-[S_{41},S_{42}]\bigr)\,, 
\label{pauli4}
\end{align}
and the projector onto the singlet states is 
\begin{equation} 
I_{j=0}=\frac{2}{3}(S_{12}S_{34}+S_{13}S_{24}+S_{14}S_{23})\,.
\end{equation}
All RFF qubit states and operators are now constructed from $I_{j=0}$ and
$\vec{\sigma}^{(\mathrm{RFF})}$ in the usual manner.  

We note that tracing out either one of the spin-1/2 constituents from the
four-qubit RFF Pauli operators in (\ref{pauli4}) leads to the three-qubit RFF
Pauli operators of (\ref{RFFqubit-3}).  
Accordingly, they are different realizations of the same su(2) Lie algebra
by representations with different dimensions.  
In other words, the amount of information carried by the logical RFF qubit is
exactly the same irrespective of the construction in terms of three or four
physical qubits. 

Further, since the Hilbert spaces for the RFF qutrit and RFF qubit are 
two terms in the direct sum of (\ref{decom}), one can utilize the two RFF
quantum systems jointly.  
The simultaneous laboratory implementation of a RFF qutrit and a RFF qubit, in
rotationally invariant states of four physical qubits, would be very
interesting indeed.

%
\section{Conclusions}
In summary, we have given the systematic and rather simple construction of RFF
quantum states in any arbitrary dimension.  
The symmetric coupling of $N$ spin-1/2 constituents distinguishes our scheme
from others.  
This novel coupling scheme immediately leads to the general explicit
expression for the RFF qudits, which also provides alternative representations
for a DF subsystem in any arbitrary dimension.  
We remark that the new expression for the DF subsystem possesses a higher
symmetry under permutations than the subsystems obtained from other
constructions, which is possibly advantageous for experimental implementations.
Future studies should address this issue.

%
\begin{acknowledgments}
JS and BGE wish to thank Hans Briegel for the hospitality
they enjoyed at the Institute for Quantum Optics and Quantum
Information in Innsbruck.
This work is supported by the National Research Foundation and Ministry of
Education, Singapore. 
Support from A*STAR Temasek Grant No. 012-104-0040, NUS Grant
WBS R-144-000-116-101, and SERC Grant 052 101 0043 is gratefully 
acknowledged.
\end{acknowledgments}

\vspace*{-1.5ex}

%

\end{document}